# Influence of magnetic field on plasma parameters and thin film deposition along axial and radial distances in DC magnetron

S. Gopikishan, I. Banerjee and S. K. Mahapatra*

*Abstract*— Magnetic field (B) distribution of the magnetron was measured and discussed its effect on plasma parameters and deposition rate. Plasma parameters such as electron temperature ($T_e$), electron number density ($n_e$) were estimated using electron flux (EF) and electron energy distribution function (EEDF) methods as function of axial, radial distances from the cathode. $T_e$ and $n_e$ decreased with increasing of axial, radial distances from the cathode. $D_r$ was obtained for the same positions where the above plasma parameters were measured, and found that similar profile with $T_e$ and $n_e$. Magnetron configuration was simulated using COMSOL Multiphysics software and compared with the experimentally measured profile. Further, density distribution was simulated using measured B through particle in cell - Monte Carlo collision and found simulation results are well supported to the experimental results.

## I. INTRODUCTION

DC magnetron sputtering is one of the admired methods for thin film deposition. In 1974 Chapins invented magnetron to improve the sputtering process [1]. Without magnetron, the system was limited to low ionization, low deposition, unstable plasma etc., [2, 3]. These limitations have been overcome by magnetron in which electrons can be trapped towards the cathode region and increase the ionization [4]. Different configurations of magnetrons like planar, cylindrical and sputtering gun designs have been used [5].

A common aspect of all these configurations is the electron trapped in the presence of electric cross magnetic field drift. Magnetron is an important unit and responsible for particle behavior in the DC magnetron plasma. Understanding of particles behavior due to magnetron in the plasma is important [6]–[13]. In recent years, studying the magnetic field distribution and its effect on plasma parameters have been gained importance to improve the quality of thin films[14]–[17]. Field distribution at axial and radial positions causes variations in plasma parameters such as electron temperature and densities[18]–[22].

S. Gopikishan and I. Banerjee, Department of Physics, Birla Institute of Technology, Mesra, Ranchi 835215, India, gopikishans@bitmesra.ac.in,indranibanerjee@bitmesra.ac.in and S. K. Mahapatra, Centre for Physical Sciences, Central University of Punjab, Bathinda, Punjab, India
skmahapatra@bitmesra.ac.in.

*Corresponding author: skmahapatra@bitmesra.ac.in.

Understanding the profile of plasma parameters will enable to optimize plasma chemistry, nucleation and growth of desired thin films. Various method such as Optical Emission Spectrometer (OES), Langmuir Probe (LP), Coherent Anti-Stokes Raman Scattering (CARS) and Laser-Induced Fluorescence (LIF) etc., have been used to measure the plasma parameters [23]–[25]. Among these methods, LP is suitable to measure low-pressure plasma parameters [26]–[30]. Intensive studies on thin film development and its characterization have been carried out over past several years [31]–[33]. J Bretagne et.al found the axial distribution of $n_e$, $T_e$ in the range of $2.5\times10^{11}$ cm$^{-3}$ to $4.5\times10^{11}$ cm$^{-3}$ and 1.5eV to 2eV respectively for Physical Vapor Deposition (PVD) magnetron sputtering system [34].

In this paper, DC magnetron sputtering system has been studied based on B distribution for the metallic thin film deposition. The quantitative analysis of plasma parameters and their contribution in deposition rate at respective axial and radial positions were analyzed. Deposited thin films were characterized by profilometer for thickness and surface morphology by field emission scanning electron microscope (FESEM). Measured magnetic field and plasma parameters were compared with COMSOL Multiphysics and particle in cell - Monte Carlo collision simulation results respectively.

## II. EXPERIMENT

The schematic diagram of the DC magnetron sputtering system is shown in Fig. 1. It is a cylindrical stainless steel chamber of diameter 270mm and length 230mm. The chamber consists of (a) DC power supply, (b) source meter, (C) Wilson-seal port, (d) top flange, (e) cylindrical shaped stainless steel chamber, (f) magnetron sputter gun, (g) target, (h) MFC gas flow meter, (i) multi Langmuir probes setup, (j) vacuum pump, (k) substrate holder, (l) bottom flange. Top flange contains sputter magnetron and multi-probes with Wilson-sealed flange, whereas, bottom flange contains the sample holder.

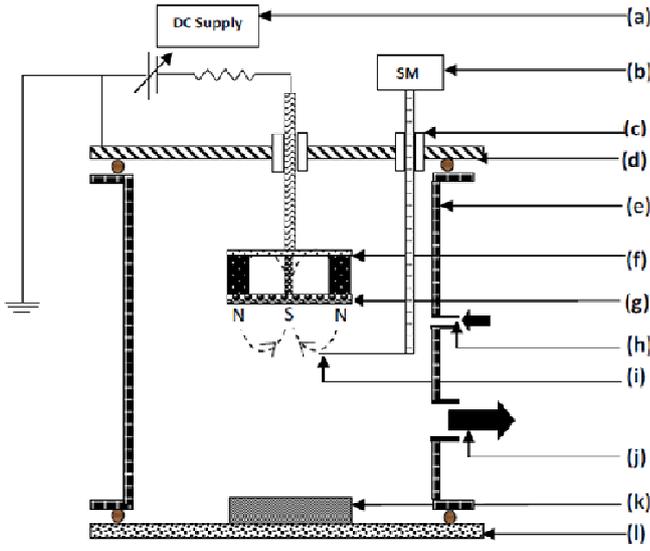

Figure. 1. Schematic diagram of a DC magnetron sputtering system

The multi Langmuir probe (MLP) setup was used in experiment and consists with four cylindrical shaped Langmuir probes, electrically isolated from each other named as A, B, C and D mounted on a quartz plate. Probe A corresponds to the central position of the cathode, whereas probe B, C and D correspond to 0.5 cm, 1 cm, and 1.5 cm away from the cathode in radial direction respectively. All probes are connected to source meter (Keithly 2410, 1100V) as shown in Fig.1. Potential from (-10 V to 10 V) were applied to MLP probes for the I-V measurement. MLP setup can move up and down without breaking vacuum pressure through Wilson-sealed port. Five axial positions are 2.5cm, 3cm, 3.5cm, 4cm and 4.5cm away from the cathode in the axial distance and named as $1^{st}$, $2^{nd}$, $3^{rd}$, $4^{th}$, and $5^{th}$ respectively. At one instant of measurement, the MLP setup was positioned at one fixed axial position and four radial positions exposed to the plasma column. The current vs voltage (I-V) plots were obtained by source meter for each Langmuir probes. The experiment was repeated for five axial positions, thus obtaining a set of 20 test data for I-V plots. Variation of $n_e$ in $m^{-3}$, $T_e$ in eV was calculated at respective axial and radial positions from the measured I-V plots.

The chamber was evacuated using a rotary pump followed by diffusion pump, with a base pressure of ~ $5 \times 10^{-6}$ mbar and working pressure ~ $6 \times 10^{-3}$ mbar controlled through Mass Flow Controller (MFC), at 20 sccm of Ar sputtering gas. Ti target (2 inch diameter and 3 mm thick) to be sputtered was clamped with the magnetron. The silicon substrates of size (5x5 $mm^2$, thickness ~0.5 mm) were cleaned by diluting HF acid, ultra-sonicated with acetone for 15 minutes and dried at room temperature prior to deposition. These Si substrates were mounted on the four different radial and five axial positions corresponding to the multi Langmuir probe positions, where the I-V data were collected using source meter (SM). During deposition, the DC power supply was fixed at 16W (320V x 0.05A). In order to estimate the deposition rate, weight of the substrate and deposited substrate were measured using Germany made analytical balance (Mettler Toeldo).The ratio of weight difference to deposition time gives deposition rate ($D_r$). For the measurement of deposition rate, MLP set was displaced and, substrates were placed at the same positions of LPs. Furthermore, thickness and surface morphology were analyzed using surface profile-meter and FESEM respectively.

### III. THEORY

#### A. Electron energy distribution function (EEDF):

Electron energy distribution function was obtained using Druyvesteyn formula [35].

$$f(\varepsilon) = \frac{4}{Ae^2}\sqrt{\frac{mV_p}{2e}} \times \frac{d^2 I_e}{dV_p^2} \qquad (1)$$

where $\frac{d^2 I_e}{dV_p^2}$ is the second derivative of the I-V plot, $V_p$ is probe voltage with respect to the plasma potential, A is the area of the probe, e and m are the electron charge and mass of the electron.

Electron density ($n_e$) and temperature ($T_e$) were calculated from the following measured distribution function,

$$n_e = \int_0^{\varepsilon_{max}} f(\varepsilon)d\varepsilon \qquad (2),$$

$$T_{eff} = \frac{2}{3n_e}\int_0^{\varepsilon_{max}} f(\varepsilon)d\varepsilon \qquad (3)$$

where $\varepsilon_{max}$ is determined from the dynamic range of the EEDF. In Druvestyn procedure, differentiation of the probe characteristics was done to determine the $T_e$ and $n_e$ corresponding to the integrals of the EEDF. The second derivative of the I-V curve gives the zero crossing point which is the plasma potential [38].

#### B. Electron flux (EF) method:

The electron temperature of the plasma was determined by measuring I–V characteristics using Langmuir probe. The slope of the ln(I) versus V curve of the I-V plot gives electron temperature ($T_e$).

$$\frac{d\ln(I)}{dV} = \frac{1}{T_e} \qquad (4)$$

where I is probe current, and V is probe voltage. Electron density obtained by using measured $T_e$ from eq[n] (4).

Electron density determined from Bohm sheath theory [36, 37] as,

$$n_e = \frac{I_{es}}{eA\sqrt{\frac{K_B T_e}{2\pi m_e}}} \qquad (5)$$

where, $I_{es}$ is electron saturation current in the I-V plot, e is electron charge, A is an area of the probe, $K_B$ is the Boltzmann constant, $T_e$ is electron temperature, and $m_e$ is electron mass. Bohm sheath theory (EF) is classical Langmuir procedure. In this method, the slope of the lnI-V curve was used to calculate $T_e$ and $n_e$. These two methods are different from each other explained in [39], [40].

## IV. RESULTS AND DISCUSSION

*A. Experimental results:*

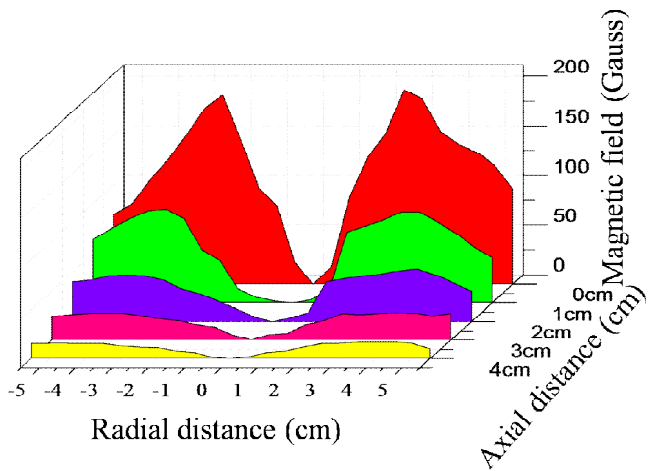

Figure. 2. Variation of magnetic field as a function of different axial and radial positions.

Magnetic field (B) distribution was measured along axial and radial distances from the surface and the center of cathode respectively using Gauss meter, depicted in Fig. 2. B is measured from the cathode surface to 5 cm away in axial outward distance and found decreasing with increasing axial distance. Radial magnetic field is symmetric from cathode center to -5cm LHS and 5cm RHS as shown in Fig. 2.

Photograph of the magnetron (Fig. 3a) and race track of the target (Fig. 3b) are shown in Fig. 3. The diameter of the magnetron is 6cm, and it consists of thirteen D63 magnets shown in Fig. 3a. In magnetron, one magnet (South Pole) at the center is surrounded by 12 magnets (North Pole). All magnets have identical strength. This magnetron (magnetic field) is clamped to the cathode and in presence of DC voltage; it creates E × B drift which aid imprisonment of electrons to the target closer region [41]. The magnetic field of the magnetron causes maximum erosion from the target known as the circular race track as shown in Fig. 3(b) [42], [43].

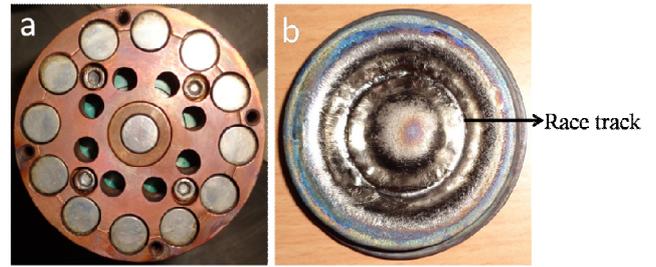

Figure . 3. (a) Magnetron, (b) Race track of the target.

Electron density ($n_e$) obtained from both EF and EEDF vs four different radial points as A, B, C and D away from the cathode center (Fig.(4I)), five different axial points are $1^{st}$, $2^{nd}$, $3^{rd}$, $4^{th}$, and $5^{th}$ away from the cathode surface (Fig. (4II)) are shown in Fig. 4. $n_e$ is decreased with increasing axial distance, away from the cathode Fig. 4(I). Figure 4(II) also shows that $n_e$ is decreased with increasing radial distance for five different axial positions.

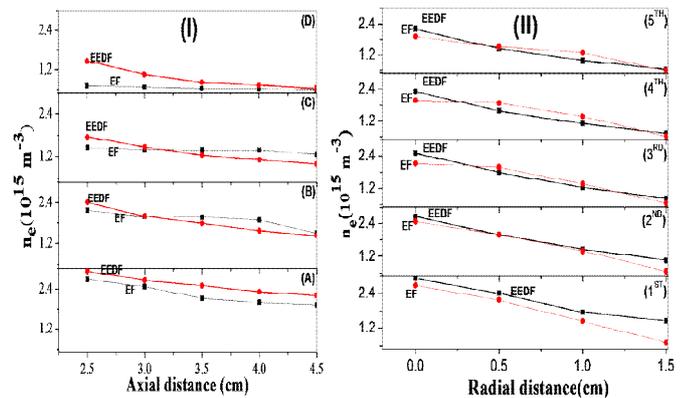

Figure. 4. Variation of electron density as a function of (I) different axial positions (II) different radial positions.

At 0cm radial distance (center of the cathode), due to presence of weaker magnetic field, electrons get entrapped, and plasma gets focused. Electrons were getting trapped from strong magnetic field to weak magnetic field. At 4.5cm axial distance electrons got diverged to a larger area, because of low magnetic field strength. Hence, number of electrons per unit volume ($n_e$) was greater at entrapped plasma. Samuel D. Ekpea also presented electron density distribution on the basis of magnetic field distribution [44].

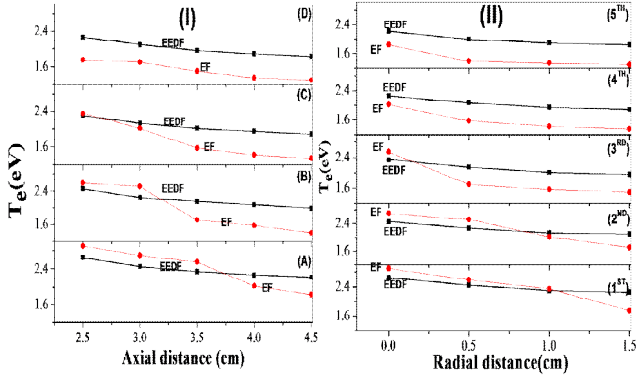

Figure. 5. Variation of electron temperature as a function of (I) different axial positions (II) different radial positions.

Electron temperature has been obtained and depicted in Fig.5 (I) and Fig. 5(II), in which $T_e$ is decreased with increasing axial and radial distances from the cathode surface and cathode center respectively. The potential was more near the magnetron; hence electrons had higher kinetic energy near the cathode and produced additional electrons in that region due to ionization by the collision. High energy electrons carry higher kinetic energy as a result of higher $T_e$ near the cathode. In axial position away from the cathode, the electrons are possibly cooled down with sputtered atoms hence possess lesser kinetic energy as a result $T_e$ decreases away from the cathode surface.

There was considerable ion loss to the greatly negative cathode, and emission of secondary electrons occurred due to ion bombardment into the plasma. These secondary electrons were accelerated by the sheath electric field and are confined by the magnetic field. Confinement of secondary electrons increases the ionization efficiency, which was the cause for high density and temperature in front of the cathode [45], [46].

The % error for $T_e$ was calculated using the equation (6) and similarly, % error for $n_e$ was calculated by replacing $n_e$ at place of $T_e$ using the following equation

$$Percentage\ error = \left|\frac{T_e(EEDF) - T_e(EF)}{T_e(EEDF)}\right| \times 100 \quad (6)$$

The % error variation for $T_e$ at corresponding axial and radial positions have been estimated i.e ~ 22% (at 2.5cm away in axial distance from cathode) to ~29% (at 4.5cm away in axial distance from cathode) and ~10% (at 0cm away in radial distance from cathode centre) to ~28% (at 1.5cm away in radial distance from cathode centre) respectively. Similarly, % error of $n_e$ is 50% (at 2.5cm away in axial distance) to ~13% (at 4.5cm away in axial distance) and ~16% (at 0cm away in radial distance) to ~50% (at 1.5cm away in radial distance).

Deposition rate ($D_r$) was obtained at the same axial and radial positions, where the $T_e$ and $n_e$ were measured. The standard eq$^n$ (7) was used to calculate deposition rate,

$$D_r = \frac{w_2 - w_1}{t_2 - t_1} \quad (7)$$

where, $w_1$ and $w_2$ are weight of silicon substrate before and after deposition, $t_1$ and $t_2$ are starting and ending time of deposition respectively. Figure 6 shows $D_r$ is followed the nature of B $n_e$ and $T_e$ correspond to radial and axial positions, is more near cathode surface than other axial & radially outward distances from the cathode. A variation in $D_r$ with magnetic field has been reported [47]–[49] which is in well agreement with the present observations.

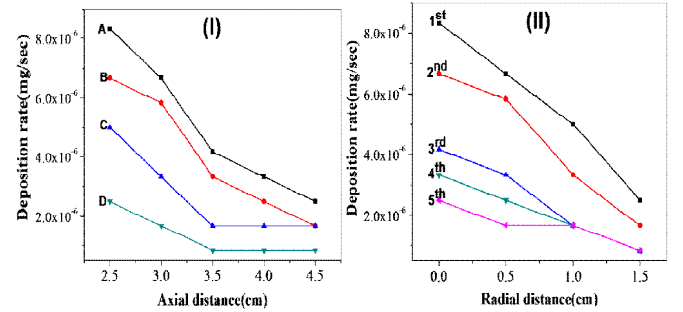

Figure. 6. Variation of deposition rate as a function of (I) different axial positions (II) different radial positions.

Surface morphologies were studied with FESEM. FESEM images of the Ti deposited thin films are depicted in Fig. 7. It was observed that with increasing axially and radially away from the cathode, decreased the density of the grains on the thin films. It was observed that the grains on the thin films also followed to the electron density profile of the plasma. At 2.5cm axial and 0cm radial distance from the cathode, the film has greater grain size compare to all other samples. Further increasing the axial and radial distances, the grain size decreased in the thin films. Compare to all the distances, thin film at 2cm radial deposition is insignificant due to confinement of plasma within the 2cm radius distance.

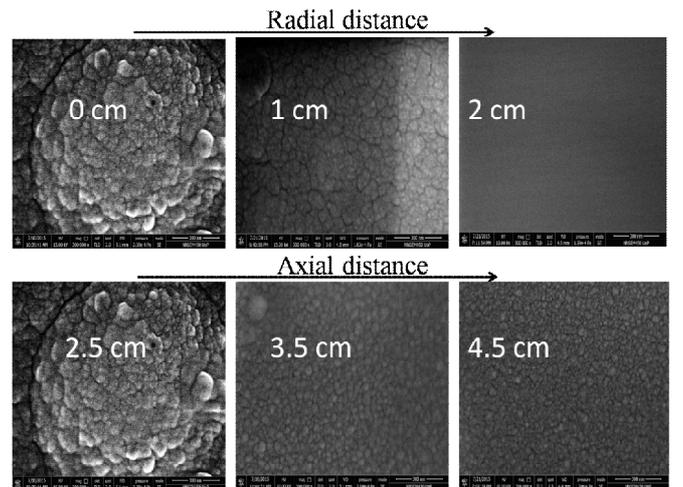

Figure. 7. Typical FESEM images of the titanium deposited thin films.

Thicknesses of the thin films are depicted in Fig. 8 and found thickness followed the plasma parameters and $D_r$.

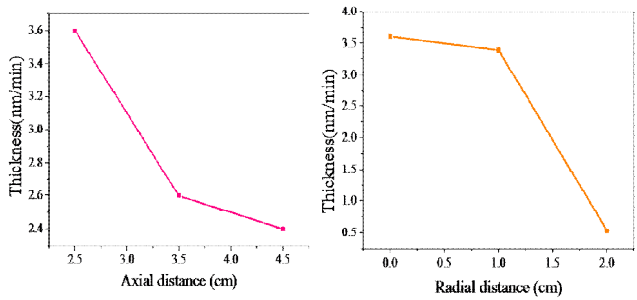

Figure. 8. Thickness of the titanium deposited thin films

B. *Simulation results:*

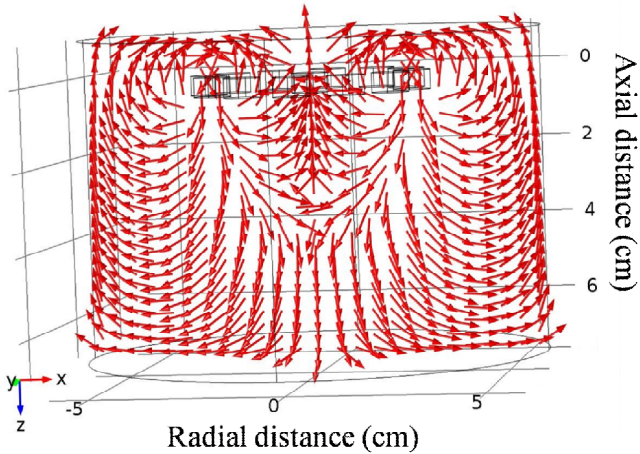

Figure. 9. Simulated results of magnetic lines of force of the present configured magnetron.

Magnetic field distribution of the magnetron was simulated using the Comsol Multiphysics 4.4 version with the aid of magnetic fields with no current (mnfc) belongs to AC/DC physics. In magnetron, one magnet (South Pole) at the center is surrounded by 12 magnets (North Pole) as shown in Fig. 3(a). Each magnet have same dimension i.e., height (h) and radius (r). Position of surrounding magnets is designed in comsol multiphysics by solving the two equations

$$x^2 + y^2 = R^2 \qquad (8)$$

$$y = \tan\theta * x \qquad (9)$$

where x,y are the coordinates of each magnet, R is radius of the magnetron and $\tan\theta$ is the slope. Magnetic lines of force obtained through simulation and shown in Fig. 9

Magnetic field distribution of the magnetron is depicted in Fig. 10. To study the variation of magnetic field distribution in axial distance, the z-coordinates has been set to 0, 2.5 and 4.5cm in multislice. Typically, variations of the magnetic field for three different distances at (a) 0, (b) 2.5 and (c) 4.5cm are shown in Fig. 10(a), 10(b) and 10(c) respectively. Fig. 10(a) shows the concentrated magnetic field whereas Fig.10(b) and 10(c) show the distributed magnetic field. It denotes that magnetic field strength is decreased in axial outward distance from the cathode. At the same plots, it is also observed that the magnetic field strength is increased in radial distance from the centre of the cathode.

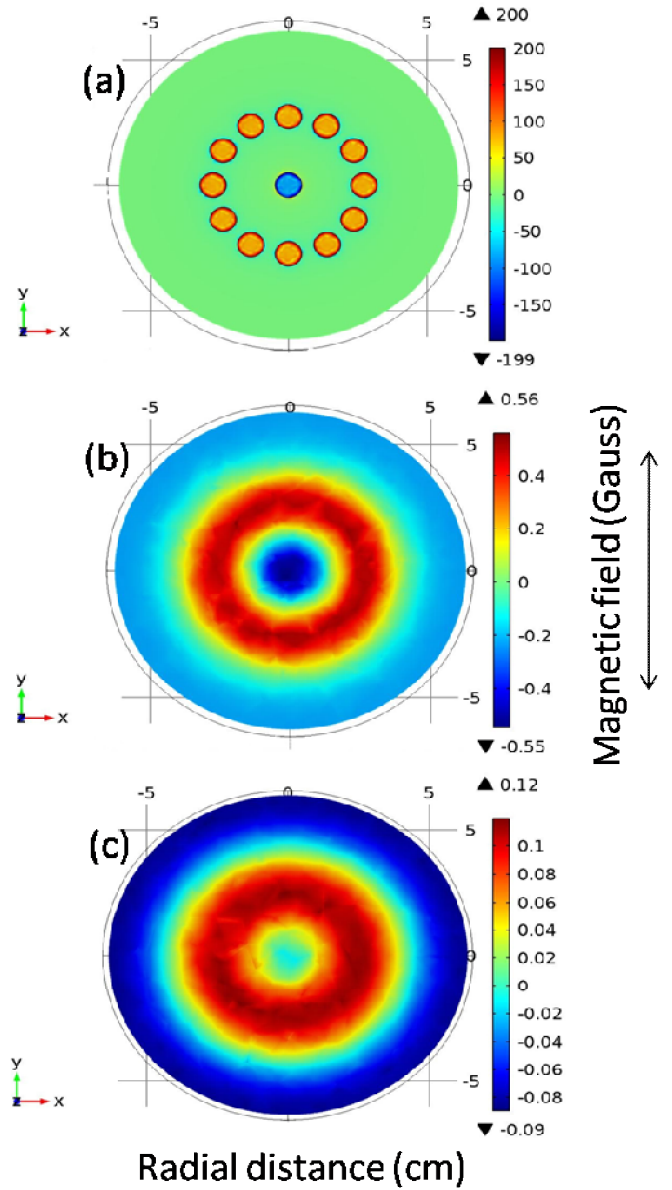

Figure. 10. Simulated result for magnetic strength of the magnetron at (a) 0cm, (b) 1cm, (c) 2cm axial distances.

The simulation of the plasma profile inside a DC magnetron was studied using Particle in Cell (PIC) approach. Simple ionization and scattering collisions were introduced into the simulation using Monte Carlo Collision (MCC) scheme. The experimental data was used for the simulation. The magnetron field was introduced at the nodes of the PIC mesh. The profile development was obtained by inserting a stable plasma into the magnetron's field with a flux velocity along the z-axis (Axial Direction). Thermal velocities along x and y directions were selected from a Maxwellian distribution of the measured temperatures.

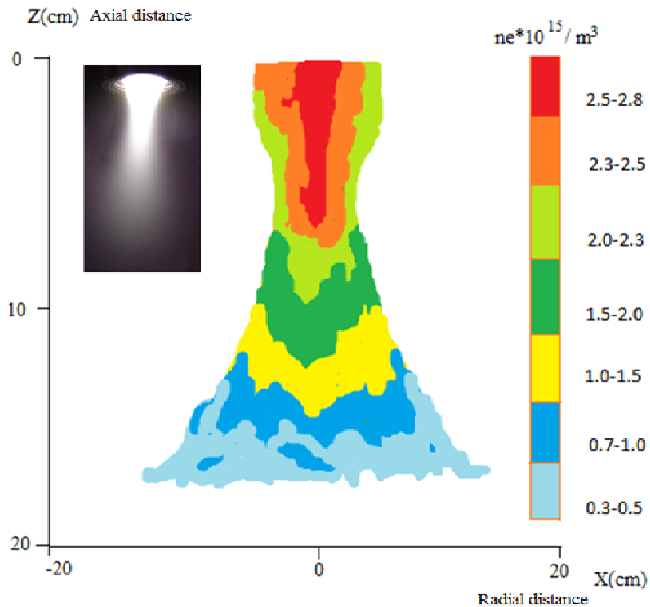

Figure. 11. Simulated density profile of the plasma for the DC magnetron plasma.

Fig.11 shows the distribution of electron density through the simulation. This is in close agreement with experimental observations (inserted figure is experimental plasma picture). Electron densities are reduced, as expected when one moves radially outwards from the cathode center and also as one moves axially down from cathode surface. Colour code profile has been used for the representation of the density profile because numerical inaccuracies of the simple PIC MCC [50], [51] code used in the present study. Limitation of the processes system could not obtain the exact simulation profile of densities at the Debye length scale. However, the profile clearly supports the essential features of the experimental findings, which has been the impetus behind the simulation attempt.

## V. Conclusion

Ions and electrons trajectory in plasma could be controlled through the magnetic field generated by configured magnetron. Plasma parameters and deposition rate at different axial and radial positions were found to be highly influenced by magnetic field distribution. Profile of the electron density in DC magnetron sputtering plasma obtained by PIC-MCC code and it has been matched with the measured profile.

ACKNOWLEDGMENT

The authors are grateful to BRNS, BARC, Mumbai, Government of India for financial support.S. G. would like to thank Stav Halder for his constant support during simulation. Authors would like to acknowledge the referees, for their suggestions to improve the quality of the paper.

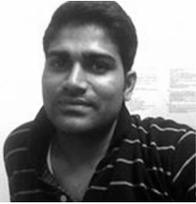

**S.Gopikishan** received the M.Sc. degree in physics with specialization of Condensed Matter Physics from the Osmania University, Hyderabad, India, in 2009, graduated from Andhra Pradesh, India, in 2007.

He is presently involved in research and development work in Investigation of Plasma Instability, BIT, Mesra. His current research interest includes nonlinear plasma dynamics. He is currently working as a Senior Research Fellow with BIT, India. He is a life member of plasma science society of India (PSSI).

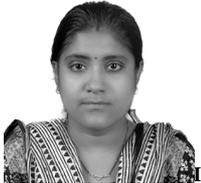

**I. Banerjee** received the Ph.D. degree on studies of emission characteristics of arc generated thermal plasma from Pune University, Pune, India in 2007.B.Sc. (Hons.) degree in physics from University of Calcutta, Calcutta, India,in 1998, and the M.Sc. degree in physics from University of Pune, Pune, India, with a specialization in materials science, in 2000.

Her field of research is plasma processing of materials. She is currently Associate Professor with Birla Institute of Technology, Ranchi, India.

.

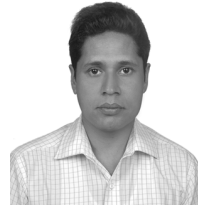

**S. K. Mahapatra** received the B.Sc. degree (Hons.) in physics from Utkal University, Bhubaneshwar, India. He received M.Sc. and Ph.D. degrees from University of Pune, India.

His fields of research are design of electron/ion accelerator, Plasma instability, Organic-inorganic solar cells and MOS/MIS devices. He is working as Associate Professor in the Centre for Physical Sciences, Central University of Punjab, Bathinda, India. Presently, He is in Brunel University of London as academic researcher for one month.